\documentclass{aa}   

\usepackage{natbib}
\bibpunct{(}{)}{;}{a}{}{,} 

\usepackage{graphicx}
\usepackage{txfonts}
\usepackage{ulem}
\usepackage{xcolor}
\usepackage{amsmath}
\usepackage{amssymb}
\usepackage{bm}

\usepackage{ulem}

\newcommand{\mat}[1]{\bm{\mathrm{#1}}}
\newcommand{\Omegam}{\Omega_\textrm{m}}

\usepackage{cancel}

\DeclareMathOperator*{\argmin}{arg\,min}

\DeclareMathOperator*{\gb} {\boldsymbol{\gamma}}
\DeclareMathOperator*{\kb}{\boldsymbol{\kappa}}
\DeclareMathOperator*{\cov} {\boldsymbol{\Sigma}}
\DeclareMathOperator*{\covn} {\boldsymbol {\Sigma_n}}

\DeclareMathOperator*{\A}{ \mathbf{A}}
\DeclareMathOperator*{\At}{ \mathbf{{\A}^{*}}}

\begin{document}

   \title{Impact of weak-lensing mass-mapping algorithms on cosmology inference}


   \author{%
    Andreas Tersenov \inst{1,2,3}\fnmsep\thanks{\email{atersenov@physics.uoc.gr}}
    \and
    Lucie Baumont
    \inst{3,4,5}
    \and
    Jean-Luc Starck
    \inst{2,3}
    \and
    Martin Kilbinger
    \inst{3}
    }

   \institute{
            Department of Physics, University of Crete, Greece
            \and
             Institutes of Computer Science and Astrophysics, Foundation for Research and Technology Hellas (FORTH), Greece
             \and
             Université Paris-Saclay, Université Paris Cité, CEA, CNRS, AIM, 91191, Gif-sur-Yvette, France
            \and
            Dipartimento di Fisica - Sezione di Astronomia, Università di Trieste, Via Tiepolo 11, 34131 Trieste, Italy
            \and
            INAF-Osservatorio Astronomico di Trieste, Via G.~B.~Tiepolo 11, 34143 Trieste, Italy
   }

   \date{Received; accepted}

  \abstract
   {
    Weak gravitational lensing is a powerful tool for probing the distribution of dark matter in the Universe. Mass-mapping algorithms, which reconstruct the convergence field from galaxy shear measurements, play a crucial role in extracting higher-order statistics from weak-lensing data to constrain cosmological parameters. However, only limited research has been done on
    whether the choice of mass-mapping algorithm affects the inference of cosmological parameters from weak-lensing higher-order statistics.
   }
   {
    This study aims to evaluate the impact of different mass-mapping algorithms on the inference of cosmological parameters measured with weak-lensing peak counts. 
   }
   {
    We employed Kaiser-Squires, inpainting Kaiser-Squires, and MCALens mass-mapping algorithms to reconstruct the convergence field from simulated weak-lensing data, generated from cosmo-SLICS simulations. Using these maps, we computed the peak counts and multi-scale wavelet peak counts as our data vectors. We performed Bayesian analysis with Markov chain Monte Carlo sampling to estimate the posterior distributions of cosmological parameters, including the matter density, amplitude of matter fluctuations, and dark energy equation of state parameter.
   }
   {
    Our results indicate that the choice of mass-mapping algorithm significantly affects the constraints on cosmological parameters, with the MCALens method improving constraints by up to 157\% compared to the standard Kaiser-Squires method. This improvement arises from MCALens’s ability to better capture small-scale structures. In contrast, inpainting Kaiser-Squires yields constraints similar to Kaiser-Squires, indicating a limited benefit from inpainting for cosmological parameter estimation with peaks.
   }
   {
    The accuracy of mass-mapping algorithms is critical for cosmological inference from weak-lensing data. Advanced algorithms like MCALens, which offer superior reconstruction of the convergence field, can substantially enhance the precision of cosmological parameter estimates. These findings underscore the importance of selecting appropriate mass-mapping techniques in weak-lensing studies to fully exploit the potential of higher-order statistics for cosmological research.
   }

   \keywords{
    statistical – gravitational lensing: weak – cosmology: large-scale structure of Universe
    }

   \maketitle
%

\section{Introduction} \label{sec:intro}

Weak gravitational lensing, or the distortion of distant galaxy images due to intervening mass along the line of sight, is a powerful cosmological probe \citep{kilbinger_cosmology_2015}.
In the past decade, weak-lensing surveys, namely the Kilo Degree Survey \citep[KiDS;][]{kuijken_KiDS_2019}, the Dark Energy Survey \citep[DES;][]{flaugher_dark_2005, DES_collab_2022}, and the Hyper Suprime Camera Survey \citep[HSC;][]{aihara_HSC_2017}, and the Ultraviolet Near Infrared Optical Northern Survey \citep[UNIONS,][]{guinot_shapepipe_2022, Li_galaxy_2024}, have imaged large areas of the extra-galactic sky to study the nature of dark matter, dark energy, and the sum of neutrino masses. 
The launch of the Euclid satellite \citep{euclid2024} heralds a new generation of so-called Stage-IV weak-lensing surveys, which will soon include the Vera C.~Rubin Legacy Survey of Space and Time \citep[LSST;][]{LSST_Large_2011}, and the Nancy Grace Roman telescope \citep{spergel_roman_2015}.  These instruments will conduct weak-lensing measurements over an unprecedented area and depth. 

Weak-lensing surveys extract cosmological information primarily by measuring two-point summary statistics: the two-point galaxy shear correlation function and its Fourier equivalent, the power spectrum.
However, current and future surveys are probing small angular scales with increasing precision.  At these scales, weak-lensing measurements contain non-Gaussian information due to non-linear structure formation that two-point statistics fail to capture \citep{weinberg_observational_2013}. 
This has motivated the development of new summary statistics, known as higher-order statistics (HOS), designed to exploit the non-Gaussian information encoded in the matter density field. Some examples include peak counts \citep{ Marian_cosmology_2009, dietrich_cosmology_2010, LK15a, LK15b, harnois-deraps_cosmic_2021, aycoberry_unions_2023, harnois-deraps_kids1000_2024}, Minkowski functionals \citep{mecke_robust_1994, kratochvil_probing_2012, grewal_minkowski_2022}, Betti numbers \citep{feldbrugge_stochastic_2019, parroni_higher-order_2021}, scattering transform coefficients \citep{cheng_new_2020, cheng_weak_2021}, three-point statistics and the bispectrum \citep{Takada_three-point_2003, takada_cosmological_2004, semboloni_weak_2011, Fu_CFHTLenS_2014}, map-level inference \citep{porqueres_lifting_2022, boruah_map_2024}, and the starlet-$\ell_1$ norm \citep{ajani_starlet_2021, ajani_starlet_2023}. \citet{ajani_forecasts_2023} performed forecasts of ten different HOS with a unified framework and found that they all outperform two-point statistics in terms of constraining power on the cosmological matter density parameter, $\Omega_\textrm{m}$, the amplitude of the matter power spectrum, $\sigma_8$, and the dark energy equation of state parameter, $w_0$. While forecasts show the very promising potential constraining power of HOS, additional studies are required to demonstrate that systematic measurement errors can be controlled for these observables in practice.

Several HOS are computed from the two-dimensional convergence, $\kappa$, which is not directly observable.  Instead, $\kappa$ must be reconstructed from galaxy shears, in a process called "mass-mapping". This reconstruction is an ill-posed inverse problem, first because of galaxy shape noise and second because of missing data from observational artefacts such as bright stars, cosmic rays, and CCD defects. As a result, the fidelity of map reconstruction differs depending on the method used (see e.g.~\citealt{jeffrey_dark_2021}). 

In the literature, metrics for comparing mass-mapping methods are typically based on the quality of the map reconstruction, with the implicit assumption that methods with a higher fidelity produce tighter cosmological contours. However, the effects of mass-mapping methods on cosmological parameters measured with HOS have not yet been fully quantified.  
Furthermore, \citet{grewal_comparing_2024}, using Minkowski functionals, found that different choices in the inversion method alter cosmological contours in a Fisher-forecast analysis.  

In this work we go a step further and measure the effect of different mass-mapping methods on cosmological constraints from peak counts in a full inference pipeline.
We focused on the following mass-mapping methods: Kaiser-Squires \citep[KS;][]{kaiser_mapping_1993}, inpainting Kaiser-Squires \citep[iKS;][]{pires_fast_2009}, and MCALens \citep{starck_weak-lensing_2021}.
KS is a linear method that is fast but implements no de-noising, and does not account for missing data or boundary effects; iKS is an improvement upon KS that provides an iteratively applied inpainting scheme to reduce the effect of missing data.
Finally, MCALens combines a sparsity-based method that captures non-Gaussian information on small scales with a Gaussian random field to replicate Gaussian behaviour on large scales \citep{starck_weak-lensing_2021}.

We examined the effects of these three inversion methods on peak counts and wavelet peak counts. These summary statistics offer several unique advantages. 
They are fast to compute and have been shown to break the degeneracy between the standard model and fifth forces in the dark sector \citep{peel_breaking_2018}.
Wavelet peak counts also have the advantages of peak counts but offer additional constraining power. In Stage-IV survey forecasts, \citet{ajani_constraining_2020} find that multi-scale wavelet peak counts yield tighter parameter constraints than a power spectrum analysis. This method is in fact so powerful that adding a power spectrum analysis does not further improve constraints compared to wavelet peak counts alone. Moreover, wavelet peak counts conveniently exhibit a covariance matrix that is nearly diagonal. 

The structure of this paper is as follows: In Sect. \ref{sec:simulations} we describe the cosmo-SLICS simulations and their associated mock data, as developed in \citet{harnois-deraps_cosmic_2021}, that are used our inference pipeline.  We review the HOS that we measure (peaks and wavelet peaks) in Sect. \ref{sec:mass_map_HOS}, which is followed by a description of our cosmological inference pipeline in Sect. \ref{sec:pipeline}. We then compare the cosmological contours derived using each mass-mapping reconstruction in Sect. \ref{sec:results}. Finally, we conclude the paper in Sect.~\ref{sec:conclusions}.

\section{Simulations} \label{sec:simulations}
One method for computing a theoretical prediction of HOS as a function of cosmological parameters is to forward-model the HOS from cosmological simulations.  In the following, we introduce the SLICS $N$-body simulations \citep{harnois-deraps_cosmic_2019}, and describe how they are used to construct mock galaxy catalogues of the DES Y1 footprint \citep{harnois-deraps_cosmic_2021}. We employed these mock galaxy catalogues to apply our mass-mapping algorithms and compute the HOS.  

\subsection{Cosmology training set} \label{sec:cosmo_sims}

In this work we used the version of cosmo-SLICS described in \cite{harnois-deraps_cosmic_2021}.
These simulations are designed for the HOS analysis of weak-lensing data, covering points over a wide range of the $\left[ \Omega_\textrm{m}, \sigma_8, h, w_0 \right]$ parameter space, where $h$ is the Hubble parameter. They sample this space at 25 $w$ cold dark matter (CDM) points as well as one $\Lambda$CDM point, arranged in a Latin Hypercube. At each of these points, a pair of $N$-body simulations with $1,536^3$ particles in a $(505 \, h^{-1}$ Mpc $)^3$ co-moving volume is run, providing 2D projections of the matter density field. 
The precise mass of each particle is determined by the simulation volume and matter density and, therefore, varies with $h$ and $\Omega_\textrm{m}$, covering the range from $1.42$ to $7.63 \times 10^9 M_{\sun}$.
The 2D projections are afterward organised into 10 (5 for each of the 2 random seeds) past light-cone density maps, $\delta_{\rm 2D} (\vec{\theta}, z)$, with a field of view of $100 \, \mathrm{deg}^2$ at different redshifts $z$.
Those maps are in turn used to obtain convergence and shear maps, $\kappa(\boldsymbol{\theta}, z_\textrm{s})$ and $\gamma(\boldsymbol{\theta}, z_\textrm{s})$, for a number of source redshifts $z_\textrm{s}$, as described in Section \ref{sec:survey_reconstruction}. 
The mean source redshift in each tomographic bin is $\langle z_\textrm{s} \rangle = 0.403, 0.560, 0.773, 0.984$.

\subsection{Covariance mocks} \label{sec:covariance_mocks}

The covariance matrix, which expresses the sample variance of our HOS, is evaluated from the SLICS suite of simulations, described in \cite{harnois-deraps_cosmological_2018}. 
This set comprises 124 fully independent $N$-body simulations, run at a fixed $\Lambda$CDM cosmology\footnote{The cosmology of the covariance training set adopts the parameters: $\Omega_\textrm{m} = 0.2905$, $\Omega_{\Lambda}=0.7095$, $\sigma_8 = 0.826$, $h = 0.6898$, and $n_\textrm{s} = 0.969$.
}, but with different random phases in the initial conditions. The specifications of this set (volume, number of particles, resolution) are the same as those of the cosmology training set. 
We further increases the effective number of simulations by a factor of 10, by generating 10 different shape noise realisations for each of the 124 simulations.  
The covariance matrix is then computed from the $1,240$ realisations of this covariance training set.

\subsection{Matching survey properties}  \label{sec:survey_reconstruction}

\citet{harnois-deraps_cosmic_2021} adapted the SLICS simulations to the DES Y1 survey by overlaying the simulated light cones with the data in a way that the resulting mock surveys have the $(\kappa, \gamma)$ values from the simulations at the positions and intrinsic ellipticities of the galaxies in the data. This was implemented by segmenting the DES Y1 catalogues into 19 tiles, each with a size of 100 $\rm deg^2$, and populating every simulated map with the galaxy samples from the corresponding tile. 
In particular, the positions of the individual observed galaxies were replicated in every simulated survey realisation. 
Every simulated galaxy was assigned a redshift by sampling from the redshift distribution $n(z)$ corresponding to its tomographic bin.

In summary, each simulated light cone from the cosmology and covariance training sets is replicated 19 times and linked to a full survey realisation.
As a final product, for the cosmology training set we have 10 mock galaxy catalogue realisations for each of the 19 tiles that cover the DES Y1 footprint, at each of the 25 $w$CDM cosmologies and the single $\Lambda$CDM cosmology. For the covariance training set, we have 124 fully independent mock galaxy catalogue realisations at the fixed cosmology, for each of the 19 tiles.

\section{Higher-order statistics on weak-lensing  mass maps} \label{sec:mass_map_HOS}

\subsection{Mass-mapping} \label{sec:mass_mapping}

The weak-lensing mass-mapping problem is the task of reconstructing the convergence field $\kappa$ from the measured shear field $\gamma$. This is possible because $\kappa$ and $\gamma$ are not independent, and can both be expressed as second-order derivatives of the lensing potential $\psi$:
\begin{equation}
    \kappa = \frac{1}{2} \nabla^2 \psi, \quad \gamma_1 = \frac{1}{2} \left( \partial_1^2 - \partial_2^2 \right) \psi, \quad \gamma_2 = \partial_1 \partial_2 \psi.
\end{equation}
where $\gamma_1$ and $\gamma_2$ are the two orthogonal components of the shear.

Rewriting the above equations in Fourier space, under the flat-sky approximation, we obtain the relation between convergence and shear:
\begin{equation}
    \tilde{\kappa}(\boldsymbol{k}) = \frac{k_1^2 - k_2^2 }{k^2} \, \tilde{\gamma}_1(\boldsymbol{k}) + \frac{2 k_1 k_2}{k^2} \, \tilde{\gamma}_2(\boldsymbol{k}), \label{gamma_to_kappa}
\end{equation}
where $\boldsymbol{k}$ is the wave vector in Fourier space.
This estimator of the convergence field, however, is only optimal in the case of zero noise, missing (or masked) data, and periodic boundary conditions.  Moreover, Eq.~\eqref{gamma_to_kappa} is only defined for $\boldsymbol{k} \neq 0$, which means that the mean $\kappa$ cannot be recovered from $\gamma$ alone. This problem is known as mass-sheet degeneracy.

To address these issues, several mass-mapping algorithms have been proposed in the literature. These techniques are based on different assumptions and methods and have been shown to produce different results in terms of the fidelity of the reconstructed maps. In this work we compared the performance of the following mass-mapping techniques: the KS method, the iterative iKS, and the MCALens method.
We provide a detailed description of these methods in Appendix~\ref{app:mass_mapping}. 

The original KS method is a direct linear inversion of the shear field to the convergence field, using the relation  Eq.~(\ref{gamma_to_kappa}). Since this method is very fast, it has been widely used in weak-lensing  studies \citep[e.g.][]{aycoberry_unions_2023, ajani_forecasts_2023, dutta_weak_2024, 2025_jeffrey_dark}. However, its simplicity leads to strong artefacts in the reconstructed convergence due to boundary effects, creates leakage between the E and B modes of the convergence field, and is not optimal in the presence of noise.
The iKS algorithm is a variation of the KS method that iteratively applies the KS inversion along with an inpainting scheme that reduces the effects of missing data. The inpainting scheme uses the morphological component analysis (MCA) iterative thresholding algorithm, introduced in \cite{starck_morphological_2005,starck_morphological_2005-1}, to fill in the missing convergence field data. This algorithm has been shown to improve the reconstruction of the convergence field in the presence of masks.
Finally, the MCALens method is a non-linear mass-mapping algorithm that combines a Gaussian component to capture large-scale behaviour and a non-Gaussian component, constructed with sparsity priors, to recover the strong peaks in the convergence field at small scales. 
As a non-linear method, MCALens is computationally more expensive, but it provides a more accurate reconstruction of the convergence field, especially at small scales, where the non-Gaussian features of the matter distribution are more pronounced \citep{starck_weak-lensing_2021}.

While there has been recent development of deep machine learning methods for mass-mapping (e.g. \citealt{jeffrey_deep_2020, remy_probabilistic_2023}), here we focus on unsupervised methods and leave comparison with deep learning methods for future work. 
Supervised methods, while promising, introduce issues such as training dataset representativity (i.e. the generalisation problem). Moreover, in terms of mean square error, deep learning techniques have been shown to achieve results similar to those of MCALens \citep{remy_probabilistic_2023}.
Additionally, although all of the above methods can be applied to data on the sphere, we instead took the approach of tiling the sky into small regions where the flat-sky approximation holds to ease comparison with the SLICS simulations during parameter inference, which are given in Cartesian coordinates.

\subsection{Higher-order statistics}\label{sec:hos}

In this section, we describe the HOS in our pipeline: peak counts and wavelet peak counts.
Weak-lensing peaks are defined as local maxima of the convergence field; in particular, we defined peaks as pixels with a value larger than that of their eight neighbours in the image. 
These peaks trace projected overdensities in the matter distribution, and thus often indicate the presence of massive cosmic structures, although the precise relation between peaks and halos is complex due to projection effects and potential noise-induced false detections \citep{sabyr_cosmological_2022}.

Peak count analysis is performed by calculating the peak function, which represents the number of peaks in a convergence map or signal-to-noise map as a function of the peak height (the respective pixel value). 
In this work we counted the peaks in signal-to-noise maps, obtained by dividing the noisy convergence maps $\kappa$, convolved with a filter $\mathcal{W}_{\theta}$, by 
its corresponding rms map, $\sigma_\textrm{n}^\mathrm{filt}$:
\begin{equation}
   S/N = \frac{\mathcal{W}_{\theta} \ast \kappa }{\sigma_{\rm n}^\mathrm{filt}} \label{snr}.
\end{equation}
Here, $\ast$ denotes the convolution operator, and ${\sigma_{\rm n}^\mathrm{filt}}$ is derived from the rms map $r$ of the shear field by
${\sigma_{\rm n}^\mathrm{filt}} = \sqrt{ \mathcal{W}_{\theta}^2 * r^2}$.

Filtering of the convergence maps is necessary to reduce noise, especially at smaller scales where noise is dominant, and to assist in the detection of peaks. 
For the filtering, we employed two different approaches: a mono-scale and a multi-scale filtering, corresponding to a single Gaussian and a set of starlet (wavelet) filters, respectively.

The mono-scale approach consists of simply convolving the convergence maps with a single Gaussian kernel of size $\theta$. 
The kernel size is chosen as a compromise between the need to reduce the noise and the need to preserve the signal.
The multi-scale peak count analysis is performed by employing a wavelet decomposition to the map, namely the isotropic un-decimated wavelet transform, commonly referred to as the starlet transform \citep{starck:sta06}. This transform enables one to extract the information encoded in different spatial scales quickly and simultaneously by decomposing the convergence maps into a set of wavelet bands of different scales. Specifically, the starlet filter decomposes the original $N\times N$ convergence map $\kappa$ into a sum of images with the same number of pixels at different resolution scales $j$:
\begin{equation}
    \kappa(x,y) = \sum_{j=1}^{j_{max}} w_j(x,y) + c_J(x,y), \label{starlet}
\end{equation}
where $c_J$ is a coarse, highly smoothed version of the original map, and $w_j$ are the wavelet bands representing the details of the original map at scale $2^j$ pixels, $j_{\rm{max}}$ is the maximum scale considered, and $J = j_{\rm{max}} + 1$. 
An important advantage of this analysis over other multi-scale approaches is that each wavelet band covers a different frequency range, resulting in a nearly diagonal peak count covariance matrix \citep{lin_quantifying_2018,ajani_higher_2021}, which does not occur, for example, in the application of a multi-scale Gaussian analysis on the convergence map.
In this starlet multi-scale approach, the standard deviation of noise $\sigma_{\rm n}^\mathrm{filt}$ must be assessed separately for each wavelet band. In our analysis, we assumed that the noise in the convergence maps is non-stationary (i.e. it depends on the number of galaxies per pixel) but Gaussian, as described in Sect. \ref{sec:catalogues}. Thanks to the linearity of the starlet transform, this pixel-level noise propagates directly to the wavelet domain, allowing us to derive a spatially dependent noise level for each wavelet coefficient at every scale and position. It has been shown \citep{starck:book15} that for starlet-decomposed images, the standard deviation of the noise at scale $j$ can be estimated from the standard deviation of the noise of the original image $\sigma_\textrm{I}$ and some coefficients $\sigma^e_j$ that represent the behaviour of the noise in wavelet space, as $\sigma_j = \sigma^e_j \sigma_\textrm{I}$. The coefficients $\sigma^e_j$ are equal to the standard deviation at scale $j$ of the starlet transform of a normal distribution with a standard deviation of one.

\section{Inference pipeline} \label{sec:pipeline}

\subsection{Shear maps} \label{sec:catalogues}

The output of the simulations described in Sect.~\ref{sec:simulations} is a set of galaxy catalogues, each containing the positions, redshifts, simulated shear components, and simulated convergence of the galaxies. We first concatenated the four redshift bins into one, so that each catalogue contained galaxies of all redshifts. We then
projected the galaxy right ascension and declination to Cartesian coordinates $(x, y)$ onto a grid with a pixel size of 1 arcmin. This resulted in $600 \times 600$ pixel maps. To create the mask, we constructed a galaxy number density map by binning the galaxy positions into the pixel grid, ($\textbf{M} = 0$ where there are no galaxies, $\textbf{M} = 1$ where there are galaxies).

Next, we binned the simulated shear components in the pixel grid to create the shear maps and generated a noise map for each shear component using 
\begin{equation}
    \sigma_\textrm{n}^2 = \frac{ \langle \sigma_{\rm int}^2 \rangle }{{2 \, n_{\rm gal}}}, \label{noise_map}
\end{equation}
where $\langle \sigma_{\rm int}^2 \rangle = 0.44$ is the average global intrinsic ellipticity dispersion of the source galaxies, and $n_{\rm gal}$ is the number of galaxies in the pixel. Here, we did not assume a constant galaxy number density but instead used the actual number of galaxies in each pixel, to create a more realistic noise map. 
This noise map was added to the shear maps to create the mock observed shear maps.

Some studies \cite[e.g.][]{marques2024cosmology, cheng2024cosmological} chose to apply some smoothing to the shear maps before the mass-mapping inversion. Our approach primarily involves filtering the reconstructed mass maps with various filters before calculating the HOS (see Section \ref{sec:hos}). However, we also incorporated a preprocessing step where the shear maps are smoothed using a Gaussian kernel with a small standard deviation of $0.4$ arcmin. This additional shear smoothing aims to reduce edge effects in the mass maps, particularly in the KS method, caused by pixels with no shear information. As shown in Section \ref{sec:mass_mapping_pipeline}, this smoothing slightly improves the reconstruction accuracy of certain mass maps. Since we subsequently convolved the reconstructed mass maps with larger filters, this initial smoothing is not expected to bias our results. Ultimately, we conducted our analysis with both smoothed and unsmoothed shear maps but find no significant difference in the resulting cosmological parameter estimation. Consequently, we report only the results obtained with shear smoothing in this paper.

\subsection{Mass-mapping}  \label{sec:mass_mapping_pipeline}

To create mass maps, we applied the KS, iKS, and MCALens methods to the noisy shear maps. For the iKS and MCALens methods, which include an internal iterative process, we set the number of iterations to 50, since this provides a good compromise between achieving sufficient convergence and managing computational time. We also set the internal parameter of the peak detection threshold in MCALens to $5\sigma$.

\begin{figure*}
    \begin{center}
    \includegraphics[width=5.0in]{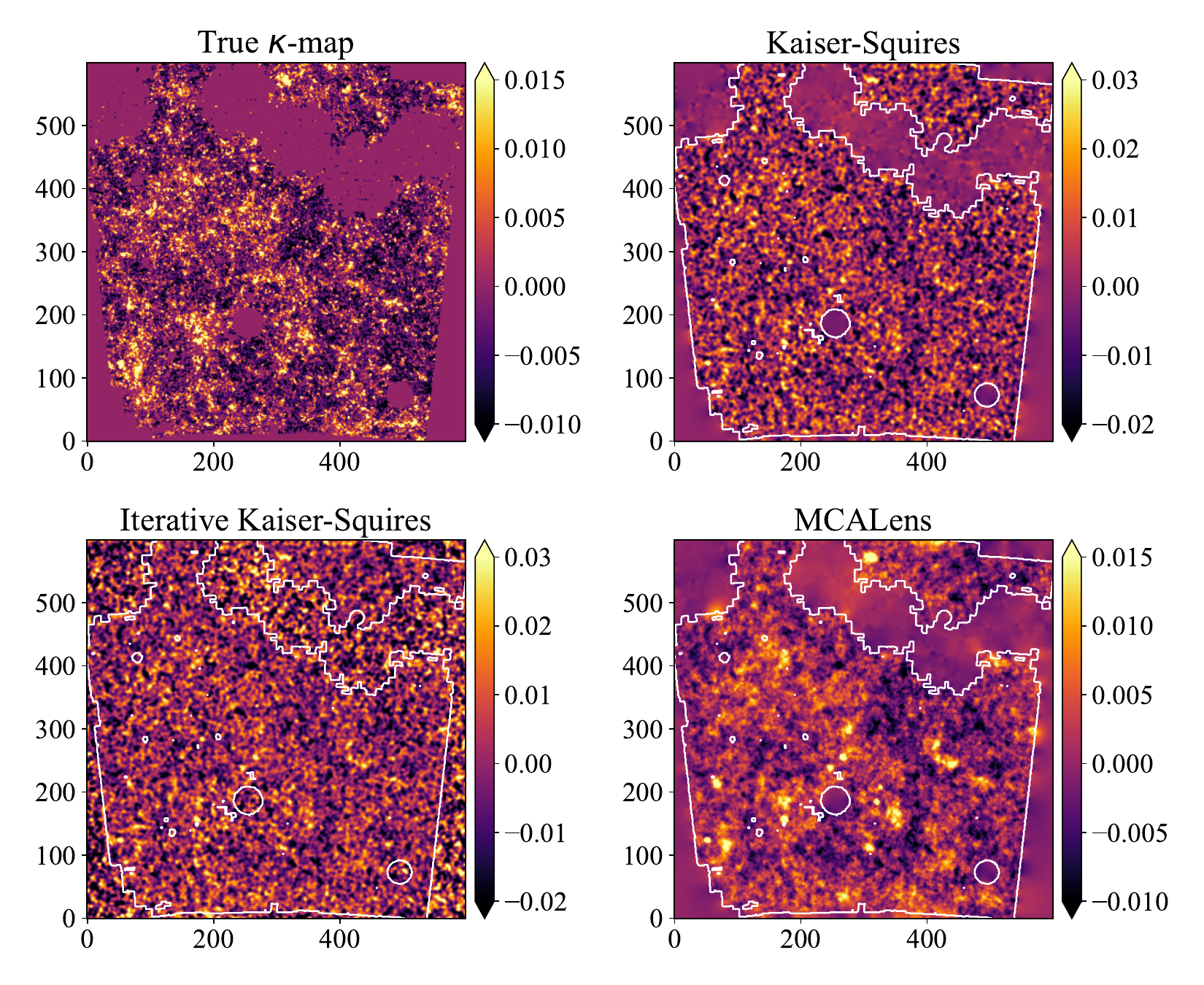}
    \caption{Mass maps obtained using different reconstruction techniques, from a specific tile of the simulated footprint. Upper left: True convergence map derived directly from the simulated $\kappa$ field (after pixelising the per-object convergence). Upper right and lower left: Mass maps generated by the KS and iKS methods, respectively, both smoothed with a Gaussian kernel (standard deviation of 2 arcmin) for visualisation purposes. Lower right: MCALens mass map. The iKS method implements inpainting of the missing data, and the unmasked regions of the patch are outlined in white. Note that the white line outlines only the large unmasked areas, for better visualisation purposes.}
    \label{fig:mass_maps}
    \end{center}
\end{figure*}

In Figure \ref{fig:mass_maps}, we show the mass maps obtained using the KS, iKS, and MCALens methods from a single tile of the simulated footprint. The KS and iKS maps have been smoothed with a Gaussian kernel with a standard deviation of $2$ arcmin, to render the structures more visible. Without smoothing, the KS and iKS maps are dominated by noise. The MCALens maps, however, are already noise-suppressed by the mass-mapping algorithm, so they do not require additional smoothing for visualisation purposes.

The iKS map uses inpainting to fill in the missing data in the convergence field (see Sect. \ref{sec:mass_mapping}), effectively reducing the edge effects visible in the KS map. For reference, large masked regions are outlined in white in the plots of the reconstructed mass maps. 
Additionally, we see that the MCALens method removes noise, while keeping the stronger peaks in the convergence field, as expected from the method.

Table \ref{tab:mass_map_stats} presents a quantitative comparison of the reconstructed mass maps generated using the KS, iKS, and MCALens methods. The comparison evaluates shear maps processed with or without Gaussian smoothing (standard deviation $0.4$ arcmin). The reconstruction accuracy is measured using the root mean square error (RMSE) between the true convergence map, $\kappa_{\rm true}$, and the reconstructed map, $\kappa_{\rm rec}$, calculated as:
\begin{equation}
\textrm{RMSE} = \sqrt{ \frac{1}{N_{\rm mask}} \sum_{i} \mathbf{M}_i \left( \kappa_{\rm true, i} - \kappa_{\rm rec, i} \right)^2 },
\end{equation}
where $i$ is the pixel index, $N_{\rm mask}$ is the total number of unmasked pixels (i.e., where $\mathbf{M}_i = 1$), and $\mathbf{M}$ is the binary mask. The RMSE values reported in the table are computed as the mean over all independent realisations and all cosmologies. The table also lists the error on this mean RMSE. As shown in the table, applying Gaussian smoothing to the shear maps reduces the RMSE for the KS and iKS methods, indicating improved reconstruction accuracy. However, shear smoothing has minimal effect on the MCALens method, which consistently outperforms the others by achieving the lowest RMSE.

\begin{table}
\caption{RMSE comparison for mass map reconstruction}
\label{tab:mass_map_stats}
\centering
\begin{tabular}{ll}
\hline\hline
Method                           & RMSE $\downarrow$ \\ 
\hline
KS (no $\gamma$ smoothing)       & $(1085 \pm 2) \times 10^{-5}$ \\
iKS (no $\gamma$ smoothing)      & $(1100 \pm 2) \times 10^{-5}$ \\
KS (with $\gamma$ smoothing)     & $(1018 \pm 2) \times 10^{-5}$ \\
iKS (with $\gamma$ smoothing)    & $(1023 \pm 2) \times 10^{-5}$ \\
MCALens (no $\gamma$ smoothing)  & $(978 \pm 2) \times 10^{-5}$ \\
MCALens (with $\gamma$ smoothing)& $(976 \pm 2) \times 10^{-5}$ \\
\hline
\end{tabular}
\tablefoot{
The RMSE is evaluated between the true convergence map and the reconstructed mass maps obtained via the KS, iKS, and MCALens methods. It is evaluated for mass maps derived from shear maps that are either smoothed with a Gaussian kernel of standard deviation $0.4$ arcmin or left unsmoothed. The values represent the mean RMSE across all realizations of the 25 cosmologies along with the error on the mean. To ensure a fair comparison, each mass map is further smoothed with a Gaussian kernel optimized to minimize its respective RMSE.
}
\end{table}

\subsection{Statistics calculation} \label{sec:statistics}
Once the mass maps are generated, we calculated our HOS on $S/N$ maps, derived from the mass maps, as described in Section \ref{sec:hos}. We calculated the peak counts and wavelet peak counts only in regions where $\textbf{M}=1$, for both inpainted and non-inpainted maps, as the inpainted parts do not contain real signal. For both analyses, the $S/N$ maps were constructed using Eq.~\eqref{snr}, and peaks were counted and binned into 20 linearly spaced bins over the range $-2 < S/N < 6$, following \cite{ajani_constraining_2020}.

\begin{figure*}
    \begin{center}
    \includegraphics[width=8.0in]{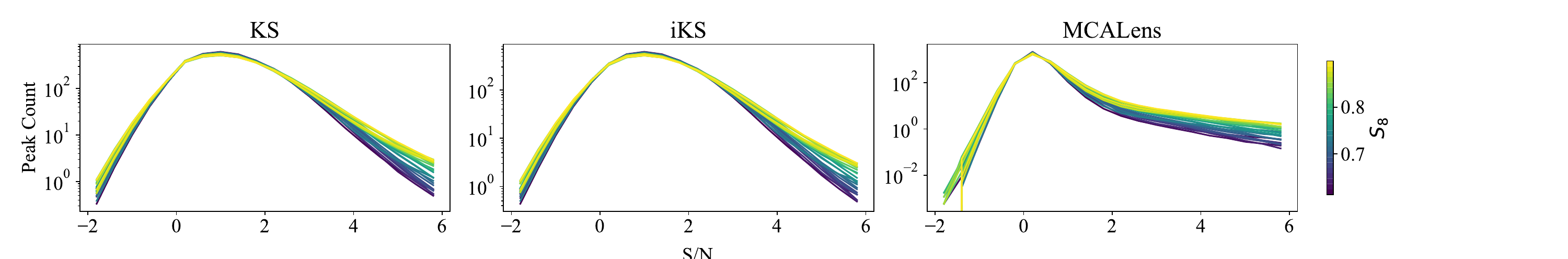}
    \caption{Log-scaled histograms of the peak counts as a function of the signal-to-noise ratio. The three panels (from left to right) correspond to the KS, iKS, and MCALens methods, for the 25 simulated cosmologies. The histograms are colour-coded by their $S_8$ value.}
    \label{fig:pc_hist}
    \end{center}
\end{figure*}

For the mono-scale peak counts, we used a Gaussian kernel as the filter $\mathcal{W}$ in Eq.~\eqref{snr}. 
We tested various kernel sizes and selected the size that yielded the best cosmological constraints for each method. The optimal kernel sizes were $\theta_{\rm ker} = 2$ arcmin for the KS and iKS methods and $\theta_{\rm ker} = 1$ arcmin for the MCALens method.
We then counted the peaks in the resulting $S/N$ maps.
Figure \ref{fig:pc_hist} displays the log-scaled histograms of the peak counts for the KS, iKS, and MCALens methods, averaged over all tiles and realisations for each of the 25 simulated cosmologies. The histograms are colour-coded as a function of the input $S_8$ showing the sensitivity of peak counts to cosmology as well as the difference between the mass-mapping methods. 
MCALens produces a significantly narrower peak distribution with a higher mode. This is a consequence of its de-noising process, which suppresses both noise and part of the signal, resulting in a lower-amplitude reconstructed $\kappa$ field compared to KS and iKS (see Fig.~1). However, the $S/N$ maps are computed using a noise level derived directly from the noise in the input shear field, Eq.~\eqref{snr}, regardless of the mass-mapping method. As a result, the reduced amplitude of the MCALens $\kappa$ map translates into a peak distribution that is more concentrated around zero.

\begin{figure*}
    \begin{center}
    \includegraphics[width=8.0in]{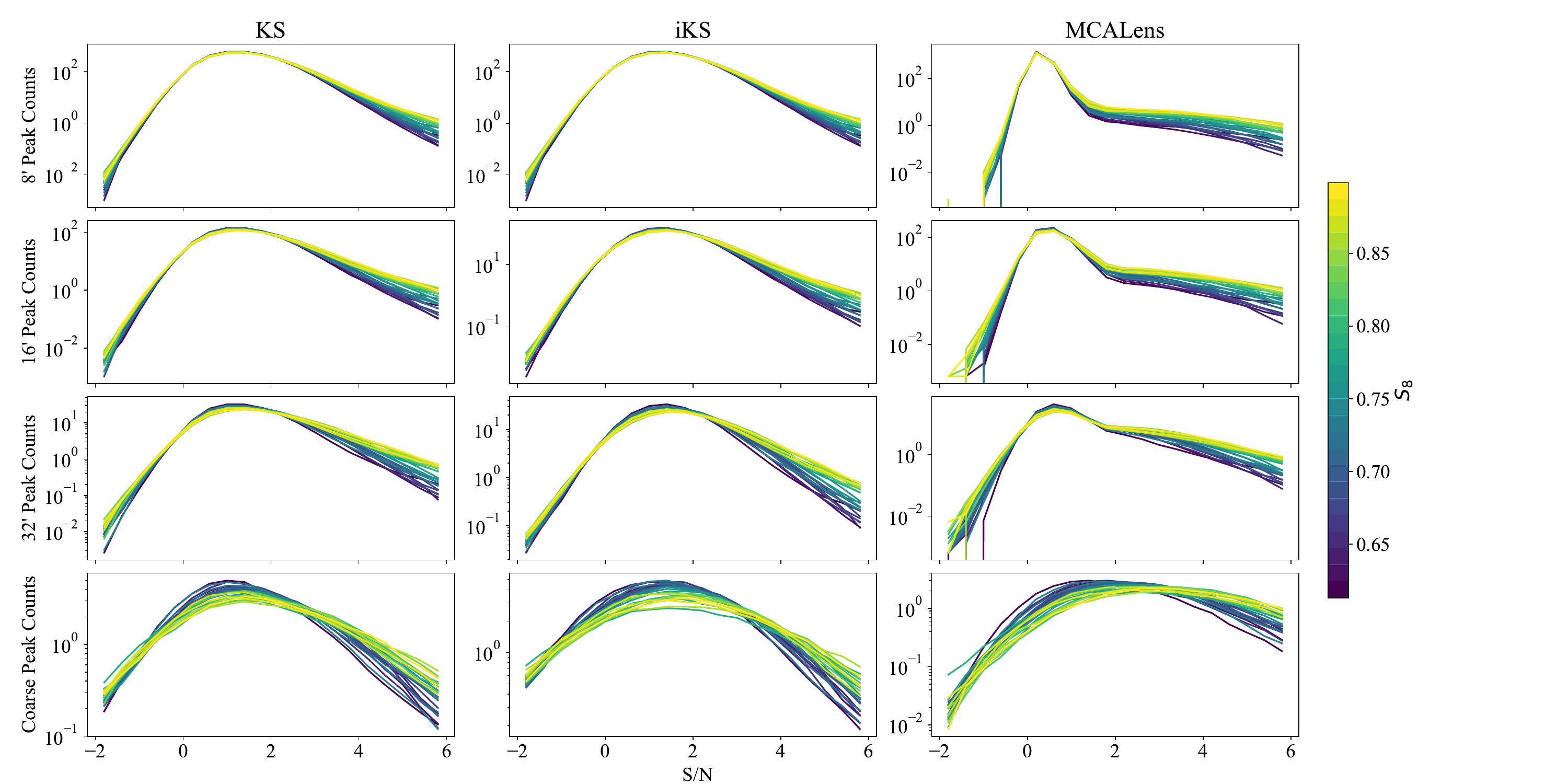}
    \caption{Log-scaled histograms of the wavelet peak counts as a function of the signal-to-noise ratio for the Kaiser-Squires (KS), iterative Kaiser-Squires (iKS), and MCALens methods, for the 25 simulated cosmologies. The four rows correspond to the three wavelet scales, $[8, 16, 32]$ arcmin and the coarse map, and the columns correspond to the three mass-mapping methods. The histograms are colour coded by their $S_8$ value.}
    \label{fig:ms_pc_hist}
    \end{center}
\end{figure*}

For the wavelet peak counts, we first decomposed the convergence maps into wavelet scales using the starlet transform, as described in Sect. \ref{sec:hos}. We used five wavelet scales, which correspond to resolutions $[2, 4, 8, 16, 32]$ arcmin, along with the coarse map. We then calculated the $S/N$ maps for each wavelet band separately, according to Eq.~\eqref{snr}, and counted the peaks in each $S/N$ map.  
Figure \ref{fig:ms_pc_hist} shows the log-scaled histograms of the wavelet peak counts for the KS, iKS, and MCALens methods, for the 25 simulated cosmologies. As expected, scales corresponding to smaller wavelet filters contain more peaks, as they capture smaller, more abundant structures in the convergence field. As in the mono-scale case, for most scales, MCALens results in a narrower distribution of peak counts, which again is a consequence of its de-noising approach that suppresses both noise and part of the signal.

\subsection{Covariance matrices}  \label{sec:covariance}
We calculated the covariance matrices for the HOS, using the covariance training set described in Sect. \ref{sec:covariance_mocks}. To avoid unphysical large-scale mode-coupling that arises from replicating the same light cone across tiles, we randomised the selection of light cones during the survey construction, following \cite{harnois-deraps_cosmic_2021}. For each of the 124 full survey realisations, we drew the 19 tiles from different, randomly selected light cones. We applied this process to all 10 realisations and averaged them to create our final covariance matrix. The covariance matrix elements were calculated as
\begin{equation}
    {C}_{\ast, ij} = \sum_{r=1}^{N} \frac{ (x_i^r - \bar{x}_i) (x_j^r - \bar{x}_j) }{N-1}, \label{cov_matrix}
\end{equation}
where $N = 1240$ is the number of realisations in the covariance training set, $x_i^r$ is the value of the statistic in the $i$-th bin of the $r$-th realisation, and $\bar{x}_i$ is the mean of the $i$-th bin over all realisations:
\begin{equation}
    \bar{x}_i = \frac{1}{N} \sum_{r=1}^{N} x_i^r.
\end{equation}

\begin{figure*}
    \begin{center}
    \includegraphics[width=5.0in]{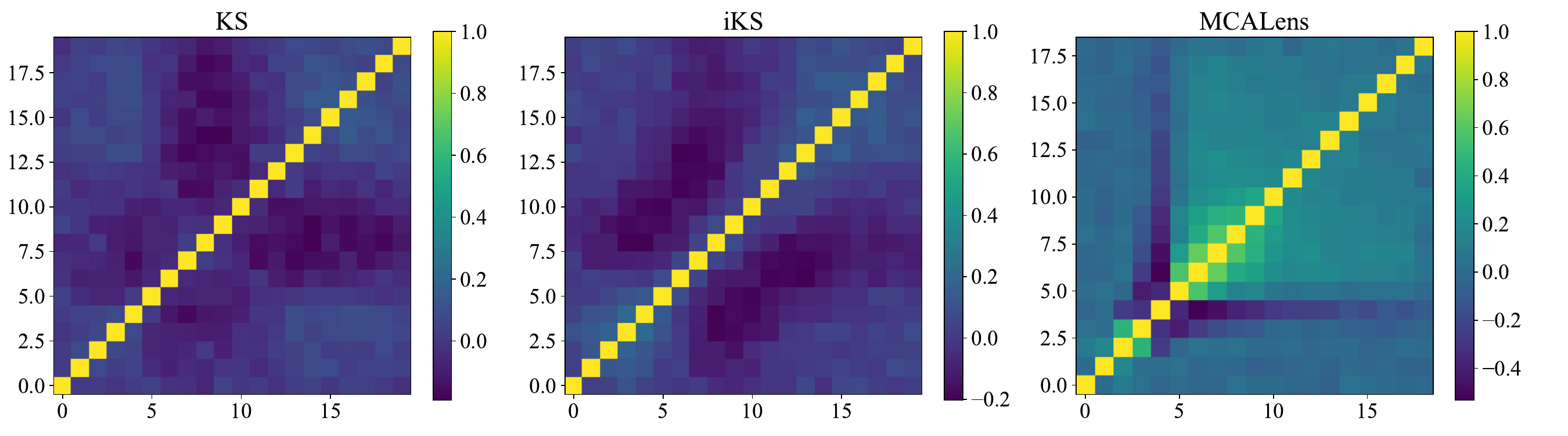}
    \caption{Correlation matrices for the single-scale peak counts, for the KS, iKS, and MCALens methods.  
    }
    \label{fig:correlation_matrices}
    \end{center}
\end{figure*}

\begin{figure*}
    \begin{center}
    \includegraphics[width=5.5in]{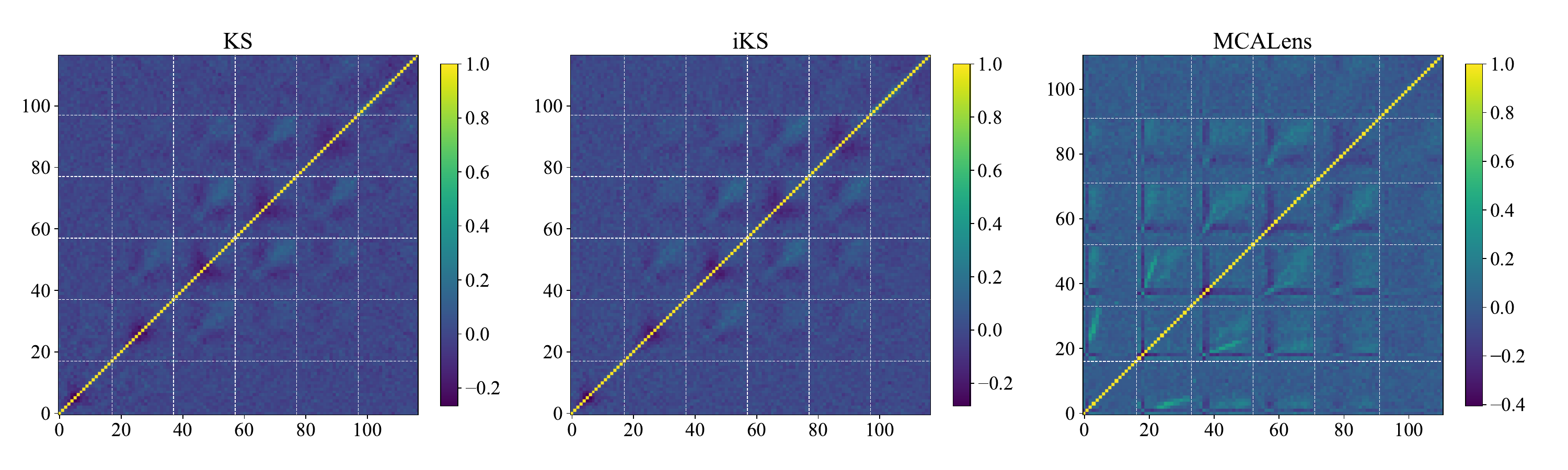} 
    \caption{Correlation matrices for the wavelet peak counts, for the KS, iKS, and MCALens methods. The dashed white lines separate the different wavelet scales, which are positioned in increasing order of resolution from left to right, i.e. $[2', 4', 8', 16', 32', \textrm{coarse}]$. 
    } 
    \label{fig:correlation_matrices_mspc}
    \end{center}
\end{figure*}

To account for the finite number of bins and realisations, we used the de-biased estimator for an inverse Wishart distribution \citep{Siskind72,hartlap_why_2007} for the inverse of the covariance matrix:
\begin{equation}
    \mat{C}^{-1} = \frac{N - n_{\rm bins} - 2}{N - 1} \; \mat{C}_*^{-1},
\end{equation}
where $n_{\rm bins}$ is the number of bins in the histogram of the HOS under consideration, and $\mat{C}_*^{-1}$ is the inverse of the respective covariance matrix calculated according to Eq.~\eqref{cov_matrix}.
In Fig.~\ref{fig:correlation_matrices}, we show the correlation matrices for the single-scale peak counts, for the different mass-mapping methods.
Fig.~\ref{fig:correlation_matrices_mspc} shows the correlation matrices for the wavelet peak counts.

\subsection{Emulator}

To obtain a prediction for our HOS for any given point in the parameter space $\left[ \Omegam, \sigma_8, h, w_0 \right]$ within the range of our simulations, we constructed an emulator using the Gaussian process regression (GPR) method \footnote{
    We used the \texttt{GPyTorch} (version 1.12) Python package \citep{gardner2018gpytorch} to implement the GPR.}.
For each bin of the HOS histograms, we trained a separate GPR model on all realisations of the HOS from the cosmology training set. 
We find that training the GPR models on the full set of realisations, rather than just the mean HOS histograms, improved the emulator's performance.
This approach enables the GPR models to predict the scaled HOS values in each histogram bin as a function of the cosmological parameters while simultaneously capturing the variability across realisations through a Gaussian likelihood.

The GPR training involves optimising the model hyperparameters by maximising the marginal log-likelihood using the Adam optimiser. We employed a radial basis function kernel with automatic relevance determination, allowing the model to learn different length scales for each cosmological parameter. To ensure stability and efficiency, the HOS values are rescaled by their mean value in each bin during training, and the scaling factors are recorded for subsequent use in predictions. 

We then validated the emulator by performing a leave-one-out cross-validation, where we trained the GPR on all but one of the 25 cosmologies in the cosmology training set, and then predicted the HOS for the left-out cosmology. 
We repeated this process for all the cosmologies in the cosmology training set, and we calculated the percentage error between the predicted and the true HOS. 
Because the training is done on one fewer cosmology than the full set, the leave-one-out cross-validation provides an upper limit on the error of the emulator.  In particular, the error is over-estimated in the case that the left-out cosmology is on the edge of the training volume, since in this case the emulator has to extrapolate from the other training points, instead of interpolating. 
We find that overall the emulator has percentage errors of less than $5\%$ for all the HOS, which is acceptable for our purposes.

However, as discussed in \cite{harnois-deraps_kids1000_2024}, the small number of available cosmologies in the cosmo-SLICS simulations used for training the emulator can lead to significant biases in the posterior distributions of the cosmological parameters. To mitigate these biases arising from the Gaussian process emulator altogether, following \cite{davies_constraining_2024}, we used the emulator's prediction of HOS at the given fiducial cosmology as our mock `observed' data vector. This choice is also made for simplicity and presentation purposes, as it assures that the contours will be centred around the fiducial cosmology, making it easier to compare the constraining power of the HOS for the different mass-mapping methods.
To ensure that this choice does not significantly alter the results of the analysis, we also perform the inference analysis using a mock data vector calculated from a cosmo-SLICS simulation at the fiducial cosmology, but in a simpler setting where the GPR is more robust. The results of this analysis are presented in Appendix \ref{sec:sim_inference}.

\subsection{Likelihood}
To perform Bayesian inference and to estimate the posterior distributions of the cosmological parameters, we model the likelihood function by assuming that our HOS follow a multivariate Gaussian distribution at each cosmology point in the parameter space:
\begin{equation}
    \log \mathcal{L} (\boldsymbol{\theta} | \boldsymbol{d}) = -\frac{1}{2} \left( \boldsymbol{d} - \boldsymbol{\mu}(\boldsymbol{\theta}) \right)^T \mat{C}^{-1} \left( \boldsymbol{d} - \boldsymbol{\mu}(\boldsymbol{\theta}) \right)  \label{likelihood}
\end{equation}
where $\boldsymbol{d}$ is the data vector of the HOS, $\boldsymbol{\mu}(\boldsymbol{\theta})$ is the vector of the predicted HOS from the emulator at the point $\boldsymbol{\theta}$ in the parameter space, and $\mat{C}$ is the covariance matrix of the HOS.

\subsection{MCMC and posterior distributions}

We performed a Markov chain Monte Carlo (MCMC) analysis employing the \texttt{emcee}\footnote{https://github.com/dfm/emcee} Python package. We assumed flat priors for the parameters $\Omega_\textrm{m} \in [0.1019, 0.5482]$, $\sigma_8 \in [0.4716, 1.3428]$, $h \in [0.6034, 0.8129]$, and $w_0 \in [-1.9866, -0.5223]$, which are the boundaries of the training volume. We used the Gaussian likelihood function in Eq.~\eqref{likelihood} and a model-independent covariance, as described in Sect. \ref{sec:covariance}. To verify the convergence of the chains, we used the Gelman-Rubin diagnostics
\citep{gelman_inference_1992}.

\section{Results} \label{sec:results}

In this section we present the results of our inference analysis for the parameters $\Omega_\textrm{m}$, $\sigma_8$, $h$, and $w_0$, using the pipeline described in the previous section. We show the parameter constraints, estimated from peak counts and wavelet peak counts, and we investigate the impact of the choice of the mass-mapping method on these constraints.

\subsection{Mono-scale peak counts}

\begin{figure}[h]
    \begin{center}
    \includegraphics[width=3.3in]{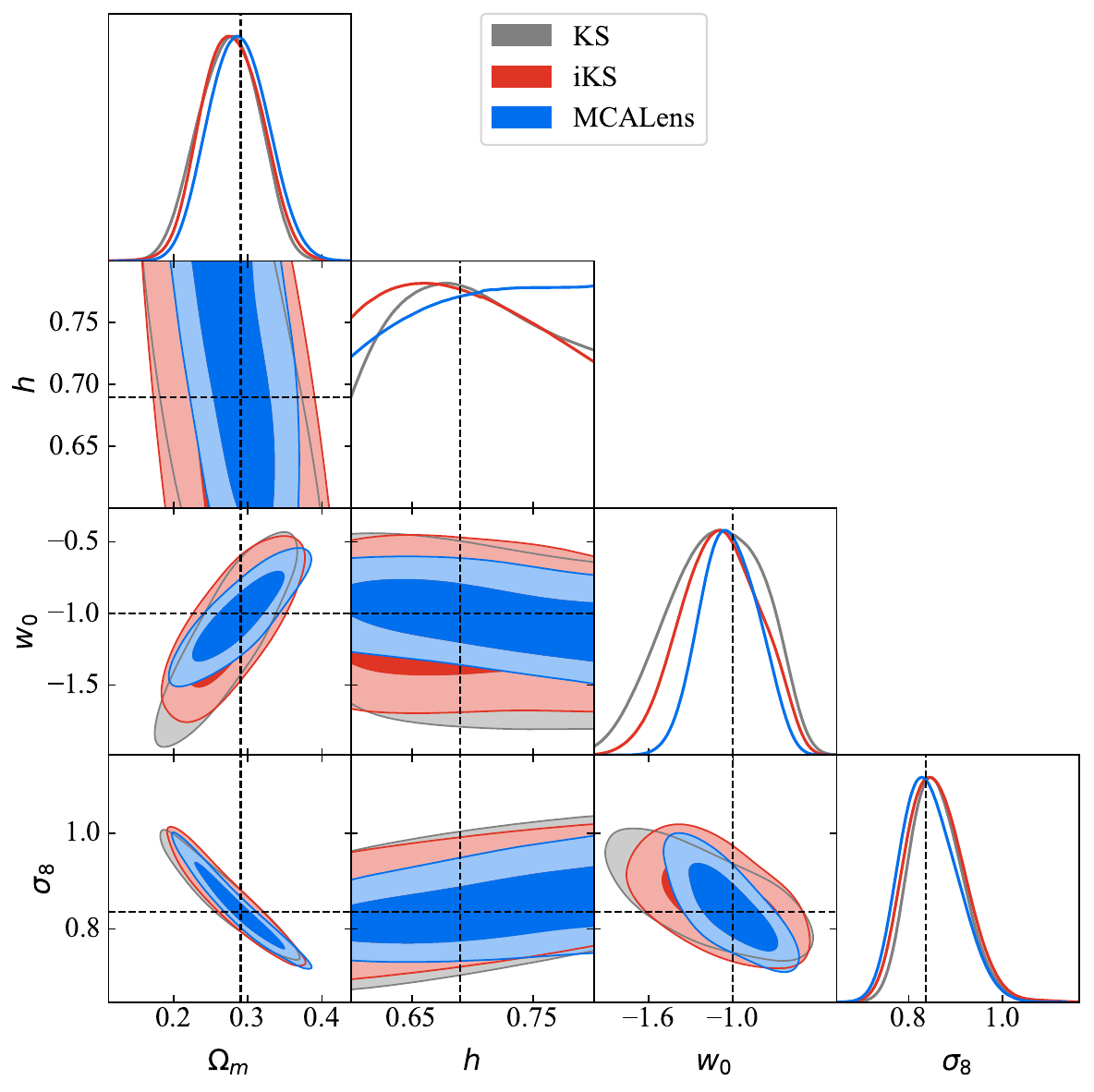}
    \caption{Constraints on the parameters $\Omegam$, $\sigma_8$, $h$, and $w_0$ from the single-scale peak counts, for the KS (grey), iKS (red), and MCALens (blue) methods. The contours represent the $68\%$ and $95\%$ confidence levels, and the dashed black lines indicate the true parameter values of our fiducial ($\Lambda$CDM) cosmology.}
    \label{fig:ss_pc_constraints}
    \end{center}
\end{figure}

Figure \ref{fig:ss_pc_constraints} shows the constraints obtained from single-scale peak counts for the KS,  iKS, and MCALens methods. We observe a substantial difference in the constraints obtained from the different mass-mapping methods, with the MCALens method providing the tightest constraints, followed by the KS and the iKS method. 

We further quantified these results by calculating the figure of merit (FoM) for each pair of parameters, defined as
\begin{equation}
    \textrm{FoM} = \left( \det \tilde{F} \right)^{1/n},
\end{equation}
where $\tilde{F}$ is the marginalised Fisher matrix, estimated as the inverse of the covariance matrix among the parameters of interest, calculated from our MCMC chains, and $n$ is the number of parameters. The FoM is a measure of the constraining power of the data on the parameters, with higher values indicating tighter constraints.
Table \ref{sspc_fom_tab}
shows the FoM for each subset of parameters, as well as for the full set of parameters, for the KS, iKS, and MCALens methods. 

We can see that using MCALens instead of KS improves the FoM by 17$\%$. This improvement can be attributed primarily to the de-noising strategy employed by MCALens. Specifically, MCALens's wavelet-based de-noising technique is more effective at preserving true peaks at small scales, which are critical for constraining cosmology. In contrast, KS and iKS rely more on Gaussian smoothing, which reduces noise but sacrifices some of the small-scale signal. Moreover, iKS does not provide a clear improvement over KS, sometimes performing slightly worse than KS. This is likely because the inpainting applied by iKS primarily affects the masked areas, which are excluded from the peak count analysis. This may also indicate that the optimal Gaussian smoothing for iKS was not fully reached with the kernel sizes we explored, leading to suboptimal performance with our current choices. We note that for other HOS, such as the bispectrum computed in Fourier space (which uses the full area including masked regions), the performance differences between KS and iKS could be more pronounced.

\begin{table}
\caption{FoMs for parameters from single-scale peak counts, for the KS, iKS, and MCALens methods}
\label{sspc_fom_tab}
\centering
\begin{tabular}{lccc}
\hline\hline
FoM & KS & iKS & MCALens \\
\hline\hline
$\left(\Omegam, h\right)$        & 476  & 453  & 450  \\
$\left(\Omegam, w_0\right)$       & 152  & 141  & 233  \\
$\left(\Omegam, \sigma_8\right)$  & 1323 & 1285 & 1740 \\
$\left(h, w_0\right)$            & 55   & 63   & 87   \\
$\left(h, \sigma_8\right)$       & 336  & 292  & 293  \\
$\left(w_0, \sigma_8\right)$      & 75   & 72   & 124  \\
\hline
$\left(\Omegam, h, w_0, \sigma_8\right)$ & 492 & 444 & 578 \\
\hline
\end{tabular}
\end{table}

\subsection{Wavelet peak counts}

\begin{figure}
    \begin{center}
    \includegraphics[width=3.3in]{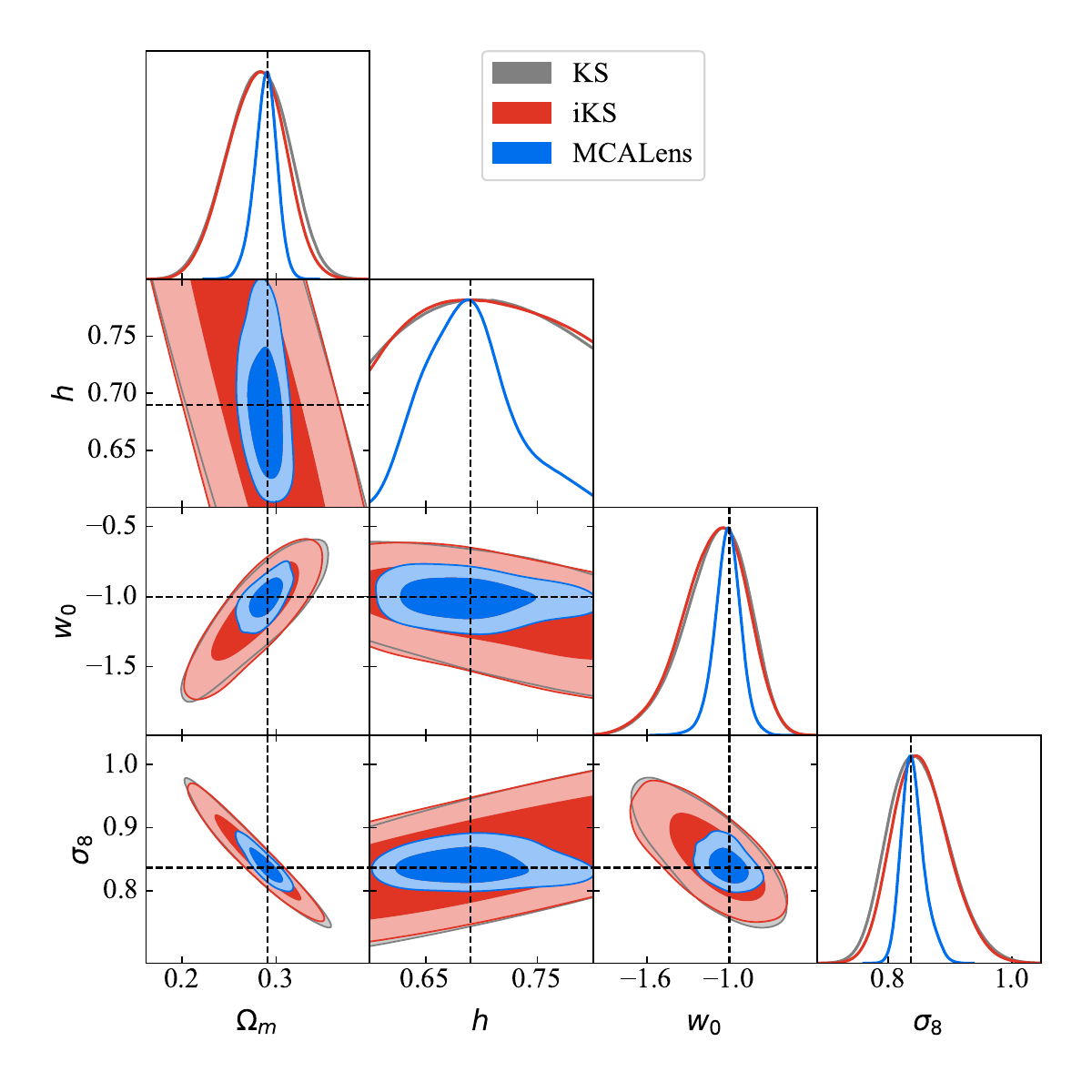}
    \caption{ Constraints on the parameters $\Omega_\textrm{m}$, $\sigma_8$, $h$, and $w_0$ from the wavelet peak counts with six wavelet scales $[2', 4', 8', 16', 32', \textrm{coarse}]$, for the KS (grey), iKS (red), and MCALens (blue) methods. The contour lines are described in Fig.~\ref{fig:ss_pc_constraints}.}
    \label{fig:ms_pc_constraints}
    \end{center}
\end{figure}

Figure \ref{fig:ms_pc_constraints} shows the constraints obtained from the wavelet peak counts, for the KS, iKS, and MCALens methods. As in the case of the single-scale peak counts, we observe a clear difference in the constraints obtained from the different mass-mapping methods, with the MCALens method delivering the tightest constraints, followed by the KS and the iKS method. Notably, KS and iKS produce very similar contours, indicating again that the inpainting applied by iKS has only a minimal effect for peak counts.
Table \ref{tab_mspc_FoM} shows the FoM for each pair of parameters, as well as for the full set of parameters, for the KS, iKS, and MCALens methods.

As expected, the constraints produced from the wavelet peak counts are tighter than those from the single-scale peak counts, across all methods. 
However, the MCALens method shows a significantly greater relative improvement in the FoM when using wavelet peak counts, with a 157$\%$ enhancement compared to the KS method with wavelet peak counts. 
Furthermore, the combination of wavelet peak counts with the MCALens method enhances the FoM by 296$\%$ compared to single-scale peak counts with the KS method.
These results underscore the critical importance of adopting a multi-scale approach, both in the mass-map reconstruction and in the statistical analysis of the convergence field.

To assess the contribution of different wavelet scales across various methods, we also performed the inference analysis incrementally by including an increasing number of scales. We began by obtaining constraints using the $[16', 32', \textrm{coarse}]$ scales. Next, we successively added finer scales to this set, starting with the $8'$ scale, followed by the $4'$ scale, and, finally the $2'$ scale.

\begin{figure}
    \begin{center}
    \includegraphics[width=3.3in]{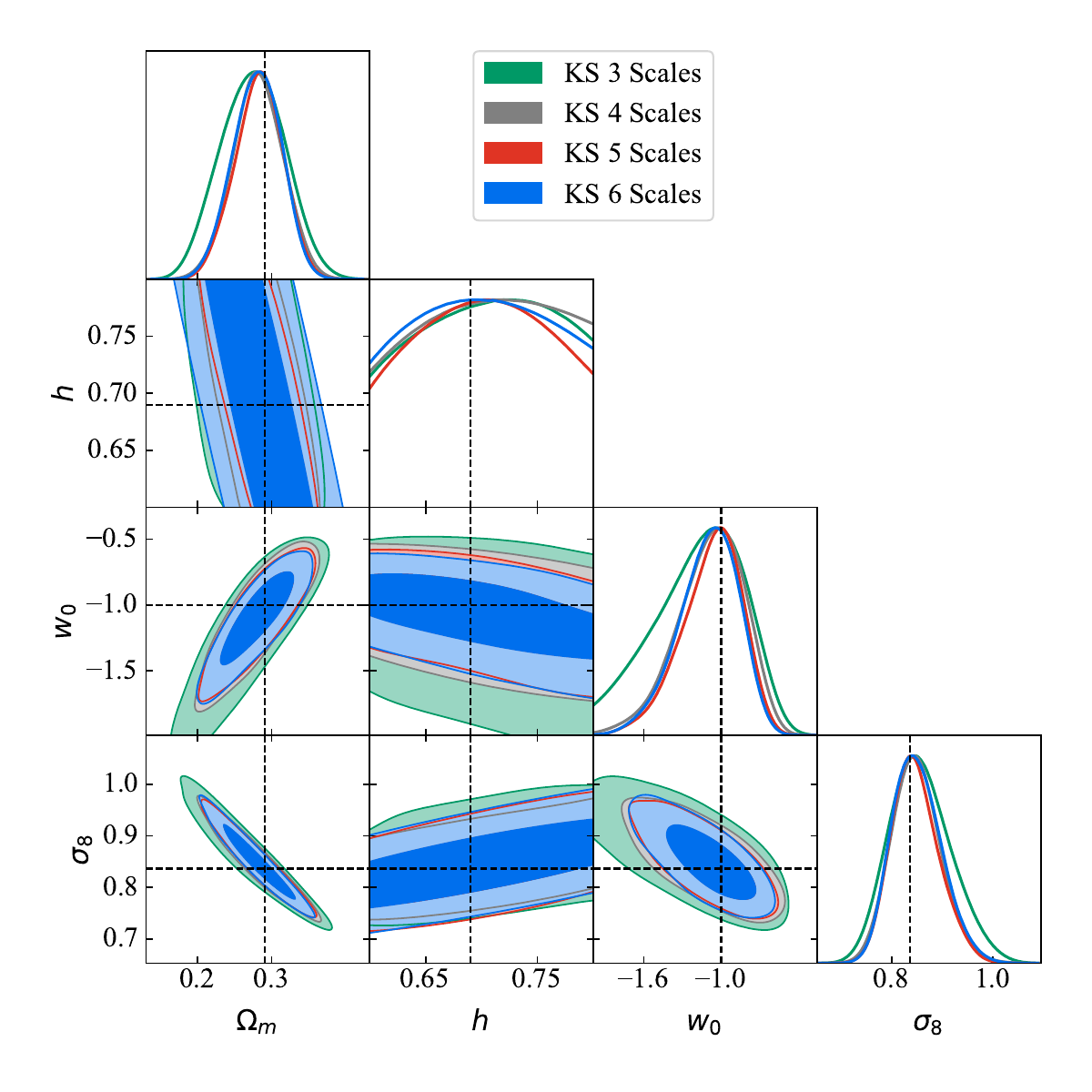}
    \caption{ Constraints on the parameters $\Omegam$, $\sigma_8$, $h$, and $w_0$ from the wavelet peak counts using KS inversion and including different numbers of wavelet scales. The green contours were obtained with three scales: $[16', 32', \textrm{coarse}]$. The grey contours were obtained by adding a fourth scale at $8'$. The red contours were obtained by adding a fifth scale at $4'$, and the blue contours include a sixth scale at $2'$. The contour lines are described in Fig.~\ref{fig:ss_pc_constraints}.}
    \label{fig:ms_pc_constraints_ks_scales}
    \end{center}
\end{figure}

\begin{figure}
    \begin{center}
    \includegraphics[width=3.3in]{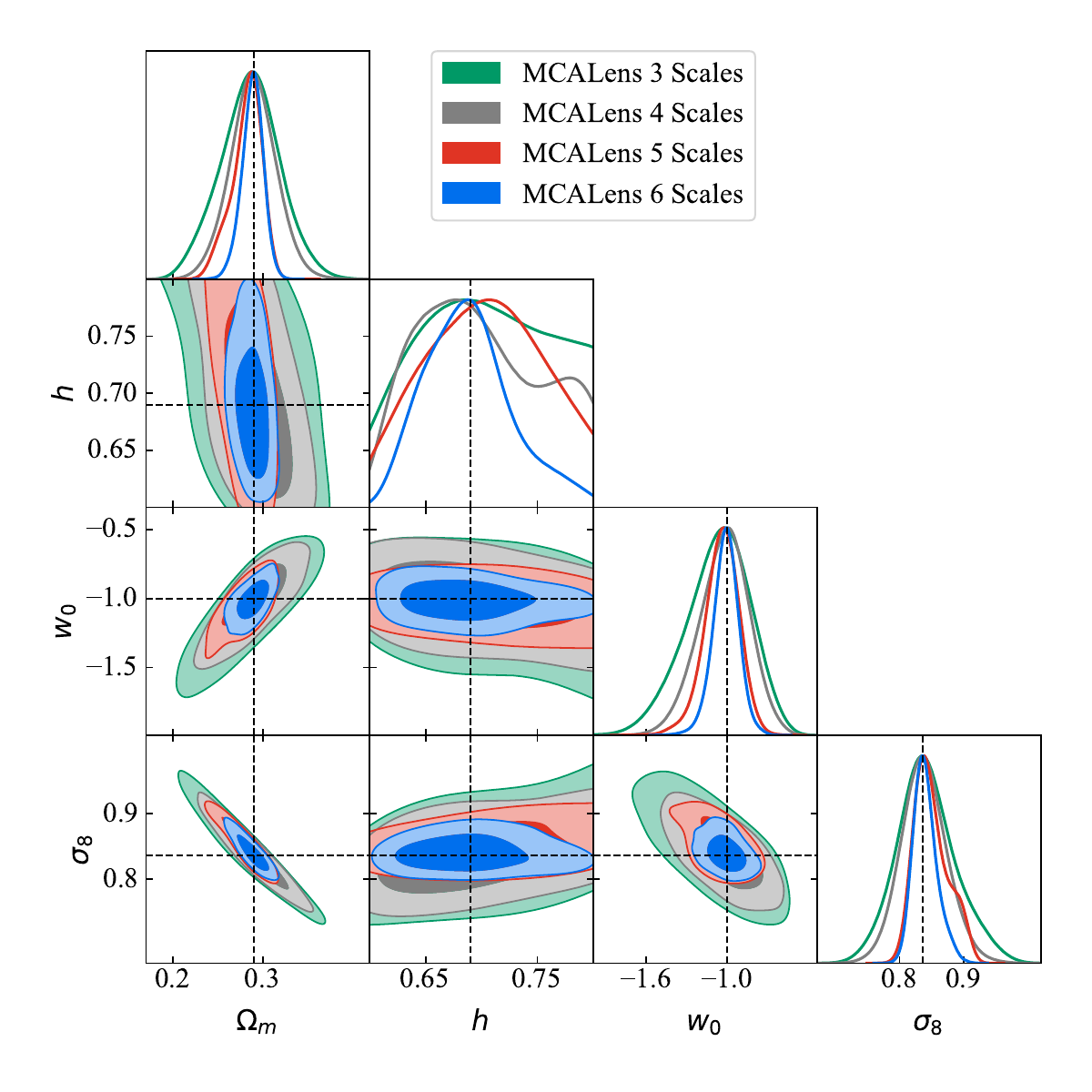}
    \caption{ Constraints on the parameters $\Omega_\textrm{m}$, $\sigma_8$, $h$, and $w_0$ from the wavelet peak counts using MCALens inversion and including different numbers of wavelet scales. The green contours were obtained with three scales: $[16', 32', \textrm{coarse}]$. The grey contours were obtained by adding a fourth scale at $8'$, the red contours a fifth scale at $4'$, and the blue contours a sixth scale at $2'$. The contour lines are described in Fig.~\ref{fig:ss_pc_constraints}.}
    \label{fig:ms_pc_constraints_mca_scales}
    \end{center}
\end{figure}

The results are shown in Figs.~\ref{fig:ms_pc_constraints_ks_scales} and \ref{fig:ms_pc_constraints_mca_scales} for the KS and MCALens methods, respectively. In Fig.~\ref{fig:ms_pc_constraints_ks_scales}, we observe that KS constraints improve significantly when including the $8'$ scale, while the $4'$ and $2'$ scales do not provide significant additional information. In contrast, Fig~\ref{fig:ms_pc_constraints_mca_scales} shows that the MCALens constraints improve significantly with the inclusion of every smaller scale. This result highlights that the MCALens method is able to extract more information from the smaller scales of the convergence field, and that is where it gains most of its constraining power over the KS and iKS methods.

\begin{table}
\caption{FoMs for parameters from wavelet peak counts, for the KS, iKS, and
MCALens methods}
\label{tab_mspc_FoM}
\centering
\begin{tabular}{lccc}
\hline\hline
FoM & KS & iKS & MCALens \\
\hline\hline
$\left(\Omegam, h\right)$        & 670  & 702  & 2159 \\
$\left(\Omegam, w_0\right)$       & 247  & 244  & 1051 \\
$\left(\Omegam, \sigma_8\right)$  & 2414 & 2517 & 9039 \\
$\left(h, w_0\right)$            & 82   & 80   & 259  \\
$\left(h, \sigma_8\right)$       & 411  & 433  & 1335 \\
$\left(w_0, \sigma_8\right)$      & 131  & 129  & 577  \\
\hline
$\left(\Omegam, h, w_0, \sigma_8\right)$ & 758 & 755 & 1947 \\
\hline
\end{tabular}
\end{table}

\section{Conclusions} \label{sec:conclusions}

We investigated the impact of various mass-mapping algorithms on cosmological inference using HOS derived from weak-lensing data. We used a set of $25$ simulated cosmologies to train an emulator for the HOS, and we used a set of $1,240$ realisations of the HOS for a fixed cosmology to estimate the covariance matrices. We then performed a Bayesian analysis with an MCMC method to estimate the posterior distributions of the parameters $\Omega_\textrm{m}$, $\sigma_8$, $h$, and $w_0$, using peak counts and wavelet peak counts as our data vectors. 

Our analysis shows that the choice of the mass-mapping method has a significant impact on the constraints on the cosmological parameters, with the MCALens method providing the tightest constraints. The improvement gained by using MCALens is $\sim 17\%$ in the FoM compared to the KS method for the mono-scale analysis, and $\sim 157 \%$ for the multi-scale analysis. We find no significant difference in the contours created with the iKS and KS methods, indicating that forward-modelling the inversion is sufficient to account for edge effects for both peak counts and multi-scale peak count statistics. However, we note that this conclusion does not necessarily generalise to all HOS. Certain HOS, such as the bispectrum, are much more sensitive to masked regions because they rely on information across Fourier modes that can be significantly impacted by missing data. In such cases, a dedicated study would be required to determine whether an inpainting approach or another treatment of the masked area is necessary to ensure reliable results. Finally, we confirm previous results, that wavelet peak counts provide tighter constraints than single-scale peak counts.  Constraints from MCALens are tightest when using a multi-scale approach because it captures additional information from smaller scales compared to KS.

These results highlight that a careful choice of mass-mapping method when attempting to accurately reconstruct the convergence field can significantly improve constraints on the cosmological parameters from HOS.

\begin{acknowledgements}
     This work was supported by the TITAN ERA Chair project (contract no. 101086741) within the Horizon Europe Framework Program of the European Commission, and the  Agence Nationale de la Recherche (ANR-22-CE31-0014-01 TOSCA). LB is supported by the PRIN 2022 project EMC2 - Euclid Mission Cluster Cosmology: unlock the full cosmological utility of the Euclid photometric cluster catalog (code no. J53D23001620006). This work used the CANDIDE computer system at the IAP supported by grants from the PNCG, CNES, and the DIM-ACAV and maintained by S. Rouberol. The authors thank Nicolas Martinet, Virginia Ajani and Joachim Harnois-Deraps for useful discussions. The authors also thank Joachim Harnois-Deraps for providing the simulations used in this work.

\end{acknowledgements}

\bibliographystyle{aa}
\bibliography{new_astro_bib}

\appendix

\section{Mass-mapping algorithms} \label{app:mass_mapping}
\subsection{Kaiser-Squires} \label{sec:ks}

In general, mass-mapping can be formulated as an inverse problem,
\begin{equation}
    \boldsymbol{\gamma} = \mathbf{A} \boldsymbol{\kappa} + \boldsymbol{n}, \label{inverse_problem}
\end{equation}
where $\mathbf{A}$ is the convolution matrix (which corresponds to the linear transformation from ideal $\kappa$-field to ideal $\gamma$-field), and $\boldsymbol{n}$ is the noise vector. The matrix $\mathbf{A}$ can be decomposed in Fourier space as $\mathbf{A} = \mathbf{F}^{*} \mathbf{P} \mathbf{F}$, where $\mathbf{F}$ is the discrete Fourier transform matrix, and $\mathbf{P}$ is the Fourier-space filter that relates the pure shear signal (in the absence of noise) to the convergence
\begin{equation}
    \tilde{\boldsymbol{\gamma}}_{\rm pure}
    = \mathbf{P} \, \tilde{\boldsymbol{\kappa}}
    = \left( \frac{k_1^2 - k_2^2 }{k^2} + \textrm{i} \frac{2 k_1 k_2}{k^2} \right) \tilde{\boldsymbol{\kappa}} .
\end{equation}
Here, $\tilde{\boldsymbol{\gamma}} = \tilde{\gamma}_1 + i \tilde{\gamma}_2$ and $\tilde{\boldsymbol{\kappa}} = \tilde{\kappa}_E + i \tilde{\kappa}_B$ are the complex representations of the two shear components and the convergence $E$ and $B$ modes, in Fourier space, respectively. 

The original method for solving the mass-mapping problem was proposed by \cite{kaiser_mapping_1993}. 
This method directly inverts the relation between the observed shear $\gamma$ and the convergence $\kappa$ in Fourier space, as given by Eq. (\ref{gamma_to_kappa}). In practice, this is done by directly applying the Fourier-space filter $\mathbf{P}^{-1}$ to the shear field:
\begin{equation}
    \tilde{\boldsymbol{\kappa}} = \mathbf{P}^{-1} \tilde{\boldsymbol{\gamma}} = \left( \frac{k_1^2 - k_2^2 }{k^2} - i \frac{2 k_1 k_2}{k^2} \right) \tilde{\boldsymbol{\gamma}}.
\end{equation}
This is the relation that was used for the standard KS reconstruction in our analysis.

While the KS method is the simplest and computationally fastest mass-mapping technique, it does not account for the noise, missing data, or the boundaries of the survey. This makes it suboptimal in the application to real data, leading to significant noise amplification, edge effects, and internal artefacts in the reconstructed maps, along with leakage between the $E$ and $B$ modes of the convergence field.
Typically, to reduce the noise in the reconstructed maps, some Gaussian smoothing is applied to the map.  This, however, also leads to a loss of information at small scales and suppression of peaks.

To rigorously account for the noise covariance matrix ${\cov}_{n}$ and the mask $\textbf{M}$, we needed to minimise the following equation:
\begin{align}
{\kb}_{ks} & = \argmin_{\kb }    \| \textbf{M}(\gamma  - \A \kb ) \|^2_{\covn},
 \label{eq:optks_inp}
\end{align}
This can be solved using the following iterative process:
\begin{align}
{\kb}_{ks}^{n+1} = {\kb}_{ks}^n   + 2 \mu \At {\cov}_{n}^{-1} (\mathbf{M}(\gb - \A  {\kb}_{ks}^n))
\end{align}
where $\mu = \min (\covn)$.
This iterative algorithm generalises the standard KS solution, to take into account the noise covariance matrix and the mask.
For the special case of a diagonal covariance matrix with all diagonal elements equal to 1 and no mask (i.e. $\textbf{M}=1$) this iterative algorithm converges to the standard KS solution. In particular, if we set $\mu = 1/2$, the update simplifies to the KS inversion in one iteration.
More details can be found in \citet{starck_weak-lensing_2021}, Appendix B.1.

\subsection{Iterative Kaiser-Squires with DCT inpainting} \label{sec:iks}

Inpainting methods are techniques used to fill in missing data in an image or signal by estimating the missing values from known data. Sparse inpainting based on the discrete cosine transform (DCT) was proposed in \citet{starck_image_2005,starck_morphological_2005} 
and used in the context of mass-mapping by \citet{pires_fast_2009} to fill in areas of missing data in the convergence field.

Inpainting can be very easily included in Eq.~\ref{eq:optks_inp}, so that one now minimises the following equation:

\begin{align}
{\kb}_{iks} & = \argmin_{\kb } \left\{  \| \textbf{M}(\gamma  - \A \kb ) \|^2_{\covn} +  \lambda  \| \boldsymbol  \Phi^{*} \kb  \|_p \right\},
 \label{eq:optw_inp}
\end{align}
 where $p=0$ or $1$, $\boldsymbol  \Phi$ is the DCT, and $\lambda$ is a Lagrangian parameter. 
The solution can be obtained using the forward-backward algorithm:
\begin{itemize}
\item Forward step:  
\begin{align}
\boldsymbol{t} = {\kb}_{iks}^n   + 2 \mu \At {\cov}_{n}^{-1} (\textbf{M}(\gb - \A  {\kb}_{iks}^n))
\label{eq:alg1_InpWiener}
\end{align}
\item Backward step:
\begin{align}
{\kb}_{iks}^{n+1} = \textbf{M}    \boldsymbol t + (1-\textbf{M})     {\boldsymbol \Delta}_{{\Phi}, \lambda} \boldsymbol t
\end{align}
\end{itemize}

where ${\boldsymbol \Delta}_{{\Phi}, \lambda}$ is the proximal operator defined in \citet{pires_fast_2009}, which consists of applying a DCT transform to $\boldsymbol{t}$, thresholding the DCT coefficients, reconstructing an image from the thresholded coefficients, and normalising the values of the filled-in pixels so that their standard deviation matches the standard deviation of the observed parts. This normalisation can also be performed at every scale of the wavelet transform of the solution. Areas for which we possess information were processed as in the iterative KS case, while the inpainting regularisation affects areas with missing data (i.e. when  $\textbf{M}=0$).

\subsection{MCALens} \label{sec:mcalens}
MCALens method was developed in \cite{starck_weak-lensing_2021}
to combine the advantages of two different methods with complementary strengths: the Wiener filter and sparsity-based methods.
The Wiener filter models the convergence field as a Gaussian random field. It effectively reconstructs the large-scale Gaussian features of the convergence field, but it is suboptimal in reconstructing the small-scale features, such as peaks. 
On the other hand, methods based on sparsity priors (e.g. \citealt{lanusse_high_2016}) assume that the convergence field is sparse when represented in a certain basis. Sparsity-based methods excel in capturing these peak structures but can struggle to retrieve the Gaussian-like statistical properties that emerge on larger scales.

To combine these two methods, MCALens models the convergence field as a sum of two components: a Gaussian component $\kappa_G$ and a non-Gaussian component $\kappa_S$, sparse in a wavelet dictionary,
\begin{equation}
    \boldsymbol{\kappa} = \boldsymbol{\kappa}_G + \boldsymbol{\kappa}_S.
\end{equation}
MCALens then estimates $\boldsymbol{\kappa}_G$ and $\boldsymbol{\kappa}_S$ by applying the MCA algorithm, which is designed to separate different components mixed within a single signal or image, provided these components have distinct morphological properties. Specifically, MCA takes advantage of the sparsity of the different components in different dictionaries, which in the case of MCALens are a weighted Fourier basis $\mathbf{\Phi}_G$ for the Gaussian component,
\begin{equation}
    \boldsymbol{\kappa}_G = \mathbf{\Phi}_G \boldsymbol{\alpha}_G = \mathbf{\Sigma}_{\boldsymbol{\kappa}}^{-1/2} \boldsymbol{\alpha}_G = \mathbf{P}_{\boldsymbol{\kappa}}^{1/2} \mathbf{F} \boldsymbol{\alpha}_G,
\end{equation}
and a wavelet basis $\mathbf{\Phi}_S$ for the sparse component,
\begin{equation}
    \boldsymbol{\kappa}_S = \mathbf{\Phi}_S \boldsymbol{\alpha}_S , 
\end{equation}  

The MCA algorithm is then solving the following optimisation problem:
\begin{equation}
    \min_{\boldsymbol{\kappa}_G, \boldsymbol{\kappa}_S} \left\{ \big\| \boldsymbol{\gamma} - \mathbf{A} \left( \boldsymbol{\kappa}_G + \boldsymbol{\kappa}_S \right) \big\|^2_{\boldsymbol{S}_n} + \mathcal{C}_G (\boldsymbol{\kappa}_G) + \mathcal{C}_S (\boldsymbol{\kappa}_S) \right\},
\end{equation}
where $\mathcal{C}_G$ and $\mathcal{C}_S$ are regularisation terms for the Gaussian and sparse components, respectively. The solution to this problem is found by iteratively applying the following steps:
\begin{enumerate}
    \item Estimate $\boldsymbol{\kappa}_G$ by assuming that $\boldsymbol{\kappa}_S$ is known, minimising:
    \begin{equation}
        \min_{\boldsymbol{\kappa}_G} \left\{ \big\| \left( \boldsymbol{\gamma} - \mathbf{A} \boldsymbol{\kappa}_S \right) - \mathbf{A} \boldsymbol{\kappa}_G \big\|^2_{\boldsymbol{S}_n} + \mathcal{C}_G (\boldsymbol{\kappa}_G) \right\}.
    \end{equation}
    \item Estimate $\boldsymbol{\kappa}_S$ by assuming that $\boldsymbol{\kappa}_G$ is known, minimising:
    \begin{equation}
        \min_{\boldsymbol{\kappa}_S} \left\{ \big\| \left( \boldsymbol{\gamma} - \mathbf{A} \boldsymbol{\kappa}_G \right) - \mathbf{A} \boldsymbol{\kappa}_S \big\|^2_{\boldsymbol{S}_n} + \mathcal{C}_S (\boldsymbol{\kappa}_S) \right\}.
    \end{equation}
\end{enumerate}

To estimate the Gaussian component, the algorithm uses the Wiener filter, performing an iteration of the proximal Wiener filtering algorithm, as described in \citet{starck_weak-lensing_2021}, Section 2, at each MCA step. The proximal Wiener filtering is a variant of the standard Wiener filter \citep{bobin_cmb_2012} which is designed to handle non-stationary noise, through the use of a Forward-Backward iterative algorithm. Thus, the penalty term $\mathcal{C}_G$ in this case is
\begin{equation}
    \mathcal{C}_G (\boldsymbol{\kappa}_G) = \left\| \boldsymbol{\kappa}_G \right\|^2_{\boldsymbol{S}_\kappa}.
\end{equation}

For the sparse component, the algorithm cannot follow the typical $\ell_1$ of $\ell_0$-norm regularisation used in the standard sparse recovery algorithms, as the large-scale wavelet coefficients overlap with low-frequency Fourier coefficients of the Gaussian field, complicating their separation. Instead, the algorithm adopts an alternative two-step approach:
\begin{enumerate}
    \item Identify the set of active wavelet coefficients by applying a threshold, typically 3-5 times the noise level at each scale $j$ and position $x$. This results in a binary mask $\Omega$, defined as:
    \begin{equation}
        \Omega_{j, x} = \begin{cases} 1 & \text{if wavelet coefficient} > \text{threshold}, 
        \\ 0 & \text{otherwise} \end{cases}.
    \end{equation}
    The mask isolates the sparse features, distinguishing them from the Gaussian component.
    \item Estimate the sparse component by solving the following minimisation problem:
        \begin{equation}
            \min_{\boldsymbol{\kappa}_S} \left\{ \big\| \Omega \odot \mathbf{\Phi}_S^\dagger \left[  \left( \boldsymbol{\gamma} - \mathbf{A} \boldsymbol{\kappa}_G \right) - \mathbf{A} \boldsymbol{\kappa}_S \right] \big\|^2_{\boldsymbol{S}_n} + \mathcal{C}_S (\boldsymbol{\kappa}_S) \right\},
        \end{equation}
        where $\odot$ denotes the element-wise multiplication, $\mathbf{\Phi}_S^\dagger$ is the adjoint operator of the wavelet transform,
        and the penalty term $\mathcal{C}_S$ is given by:
        \begin{equation}
            \mathcal{C}_S (\boldsymbol{\kappa}_S) = i_{\mathbb{R}}( \boldsymbol{\kappa}_S ) 
        \end{equation}
        where $i_{\mathbb{R}}$ is an indicator function to enforce real-valued solutions.
\end{enumerate}

This approach offers the advantage that when $\Omega$ is fixed, the algorithm behaves almost linearly, with the only non-linearity being the positivity constraint imposed on $\boldsymbol{\kappa}_S$. This enables the straightforward derivation of an approximate error map by simply 
propagating noise and relaxing the positivity constraint. 
The positivity constraint ensures that the reconstructed peaks are positive, which is not guaranteed in general, since peaks in the convergence field can be on top of voids and thus have negative values. 
The higher the non-Gaussianity of the convergence field, the more MCALens is expected to outperform linear methods like the Wiener filter.

\section{Inference with simulated observations} \label{sec:sim_inference}
In this appendix, we present the results of the inference analysis using as mock data the forward-modelled HOS from the cosmo-SLICS simulations at the fiducial ($\Lambda$CDM) cosmology. The forward model applied to generate these fiducial data vectors is the same as the one used to generate the data vectors for the emulator training, described in Sections \ref{sec:catalogues}-\ref{sec:statistics}.  

We then used the fiducial data vectors to estimate the size of the emulator's interpolation error that was not captured by our covariance matrix.  To measure this, we trained an additional GPR model on the mean values of the HOS histograms across realisations at the 25 cosmologies. From this model, we obtained the prediction uncertainties from the GPR at the fiducial cosmology. We then incorporated these interpolation errors into our error budget by adding them to the diagonal of the covariance matrix of the HOS data vectors, following \cite{heydenreich2021persistent}.

\begin{figure}
    \begin{center}
    \includegraphics[width=3.3in]{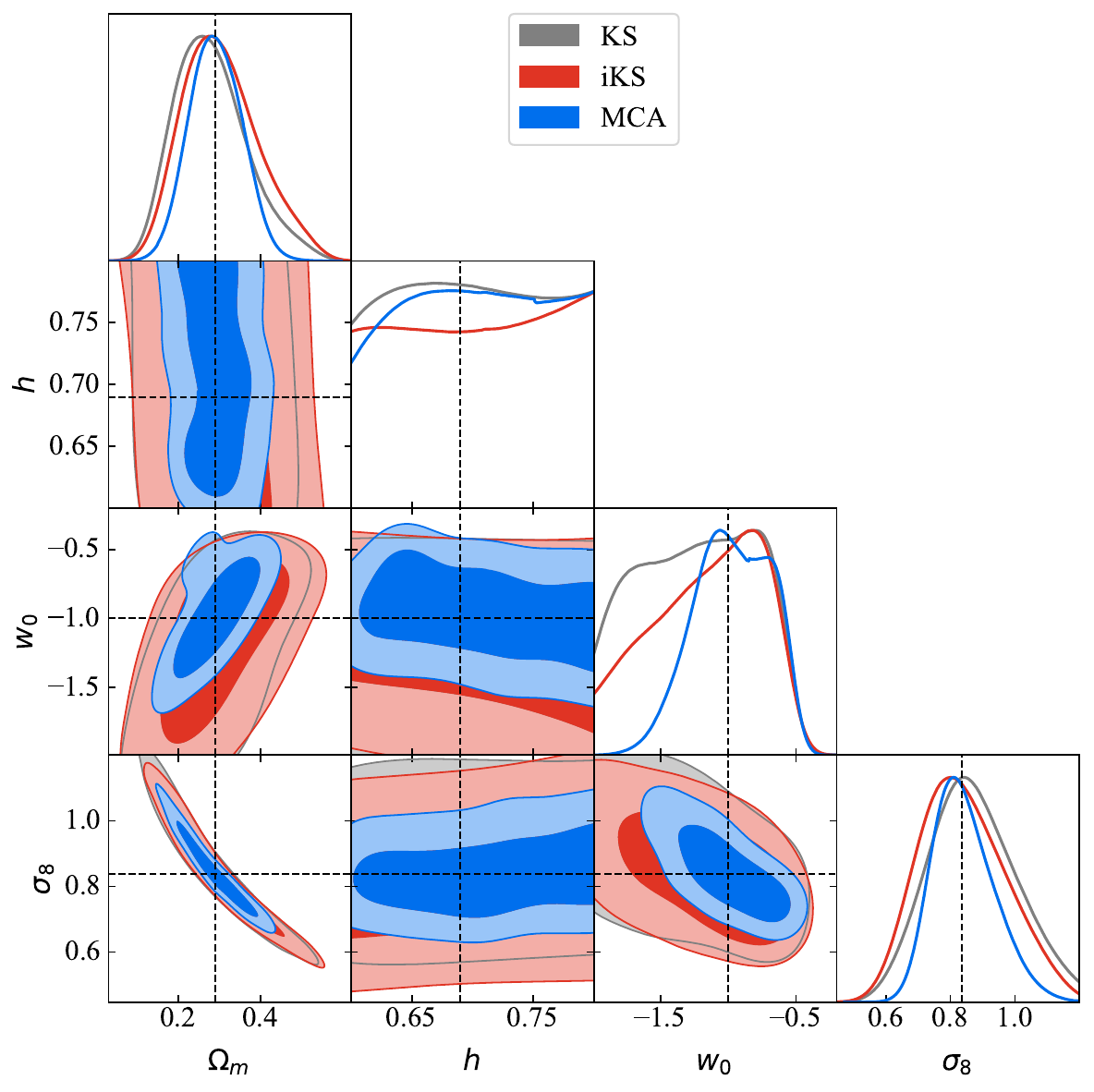}
    \caption{ Constraints on the parameters $\Omega_\textrm{m}$, $\sigma_8$, $h$, and $w_0$ from single-scale peak counts, for simulated fiducial histograms. The contour lines are described in Fig.~\ref{fig:ss_pc_constraints}.}
    \label{fig:ss_pc_sim_fid_constraints}
    \end{center}
\end{figure}

The results for single-scale peak counts are shown in Fig. \ref{fig:ss_pc_sim_fid_constraints}. The constraints on the cosmological parameters are overall larger than those obtained using the emulator predictions as mock data, shown in Fig. \ref{fig:ss_pc_constraints}, mainly due to the inclusion of the GPR interpolation error. Nevertheless, the contours still show that using MCALens instead of KS or iKS leads to significant improvements in the constraining power. 

\begin{figure}
    \begin{center}
    \includegraphics[width=3.3in]{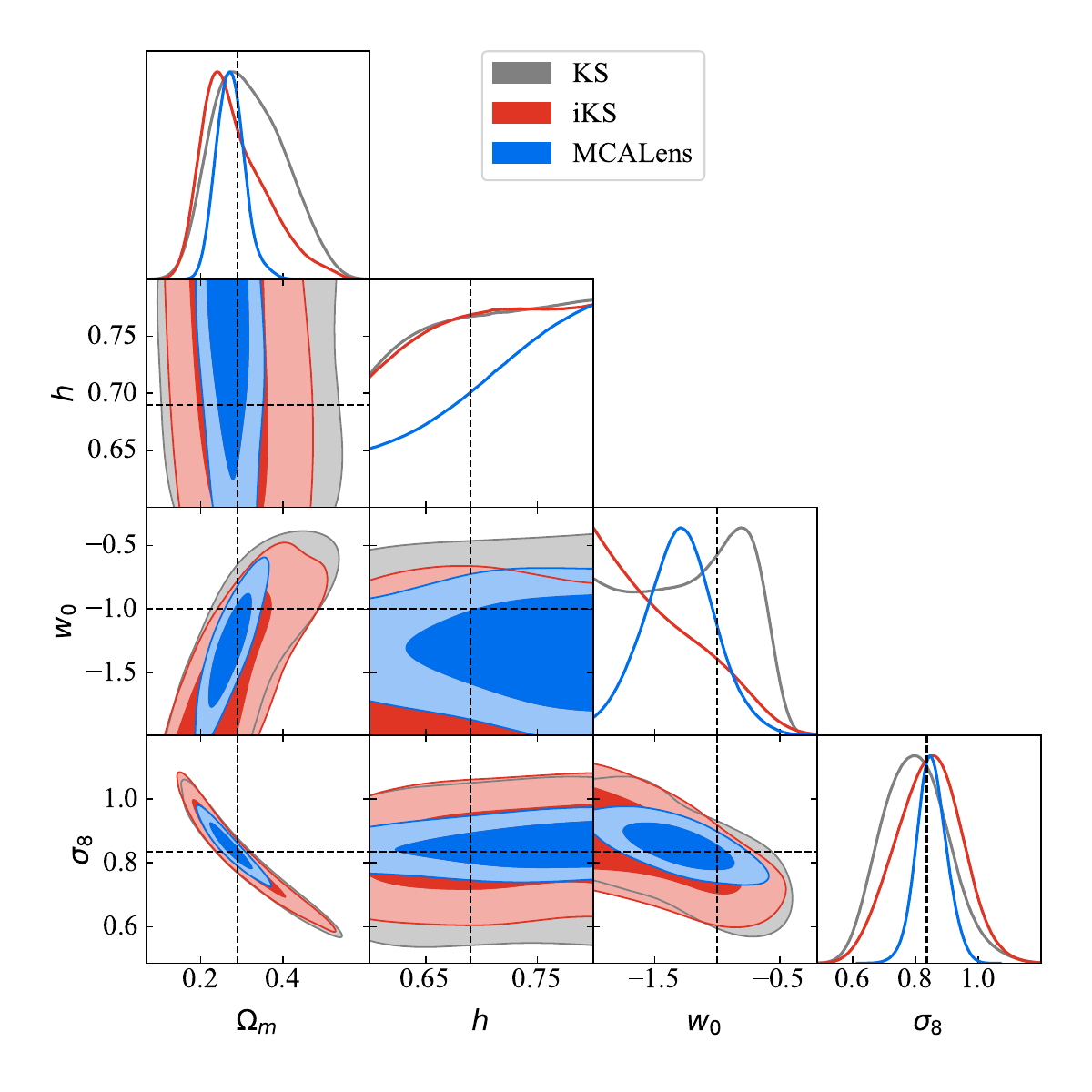}
    \caption{ Constraints on the parameters $\Omega_\textrm{m}$, $\sigma_8$, $h$, and $w_0$ from wavelet peak counts using the $[16', 32', \textrm{coarse}]$ scales, for simulated fiducial histograms. The contour lines are described in Fig.~\ref{fig:ss_pc_constraints}.}
    \label{fig:ms_pc_sim_fid_constraints}
    \end{center}
\end{figure}

The results for wavelet peak counts are shown in Fig. \ref{fig:ms_pc_sim_fid_constraints}. These constraints were obtained using only three different scales: $16', 32', \textrm{coarse}$. Including additional scales overly biased the GPR predictions due to the small number of cosmologies in the training set, leading to unreliable results. As in the case of single-scale peak counts, the resulting constraints are larger than those obtained using the emulator predictions as mock data, shown in Fig. \ref{fig:ms_pc_constraints} (both due to the use of only three scales, and the inclusion of the GPR interpolation error) but the improvement in the constraining power of MCALens over KS and iKS is still evident.

Regarding the bias in the cosmological parameters, the results show that the true values of the parameters are within the 68\% confidence intervals for all mass-mapping methods and summary statistics. However, a more detailed analysis with a larger number of cosmologies in the training set would be necessary to draw more robust conclusions. 

\end{document}